\documentclass[a4paper]{jpconf}
\usepackage{graphicx}

\begin{document}
\rightline{\small IIT-CAPP-13-11}
\title{Neutrino Factory R\&D Efforts}

\author{Daniel M. Kaplan}
\address{Illinois Institute of Technology, Chicago, IL 60616, USA}
\ead{kaplan@iit.edu}
\begin{abstract}
Stored-muon-beam neutrino factories have been recognized as the best option to measure precisely  the elements of the MNSP matrix and sensitively test the consistency of the three-neutrino mixing picture. Now that all three mixing angles have been shown to be nonzero, the motivation for neutrino factory construction is strong. A small number of feasibility issues remain open and are the subject of ongoing R\&D. Progress on these R\&D efforts is described.\\[.1in]
(Presented at NuFact 2013, 15th International Workshop on Neutrino Factories, Super Beams and Beta Beams, 19--24 August 2013, IHEP, Beijing, China.)
\end{abstract}\vspace{-.25in}
\section{Introduction}
The discovery of neutrino mixing has changed our picture of the universe and of the nature of matter and energy in important ways. It requires that the Standard Model be extended to include nonzero neutrino mass and  points the way to a deeper stratum of the underlying theory. It motivates the further study of the neutrino mixing (MNSP) matrix, with a natural benchmark of measuring its elements at least as well as those of the quark (CKM) mixing matrix. The only way to do this, the neutrino factory (NF), exploits the decay of muons\,---\,a process subject to no hadronic uncertainty\,---\,in long straight sections of a high-energy storage ring, to create well understood beams of muon and electron (anti)neutrinos that can be aimed at remote detectors~\cite{Geer,beta}.
The most poorly known elements of the MNSP matrix are $\theta_{13}$ and the {\em CP} asymmetry parameter, $\delta$. A further unknown is the neutrino mass hierarchy, i.e., which of the three known neutrino species is the heaviest. Last year $\theta_{13}$ was measured for the first time~\cite{theta13}, and it now appears likely that the mass hierarchy will be determined within the next ten years by global fits to data from near-term experiments~\cite{Blennowetal,MASS}. The {\it sine qua non} of long-baseline neutrino experimentation\,---\,the precise measurement of $\delta$\,---\,will however remain open. The experimental determination that $\theta_{13}$ is nonzero to more than five standard deviations~\cite{theta13} has been crucial to this effort, since the observability of  $\delta$ depends on all three MNSP angles being nonzero.

A series of studies~\cite{FS} have assessed the feasibility of  long-baseline neutrino factories and indicated their likely reach and cost range. A project to construct such a facility now appears practical~\cite{MASS}, once a limited set of issues have been addressed by an R\&D program now in progress~\cite{MAP}. Of these, the most challenging are the required high beam power on the pion-production target and the cooling of the muon beam in order to increase the stored intensity.

\begin{figure}
\begin{minipage}[b]{25pc}
\includegraphics[width=1.25\linewidth
]{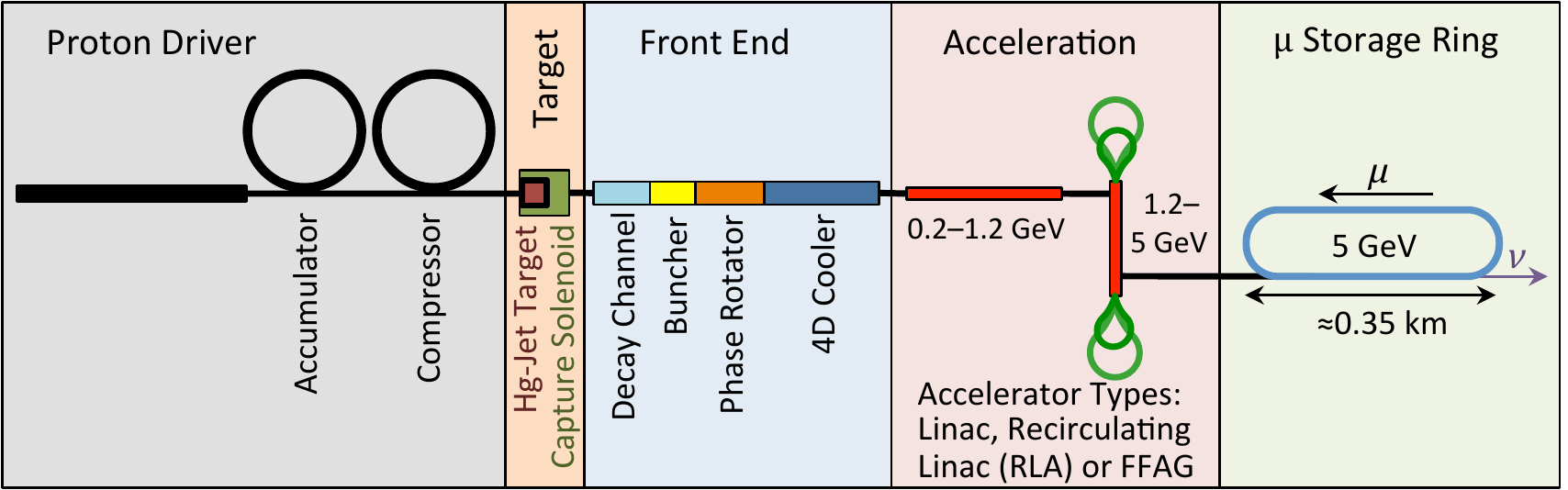}\\
\includegraphics[width=.8\linewidth
]{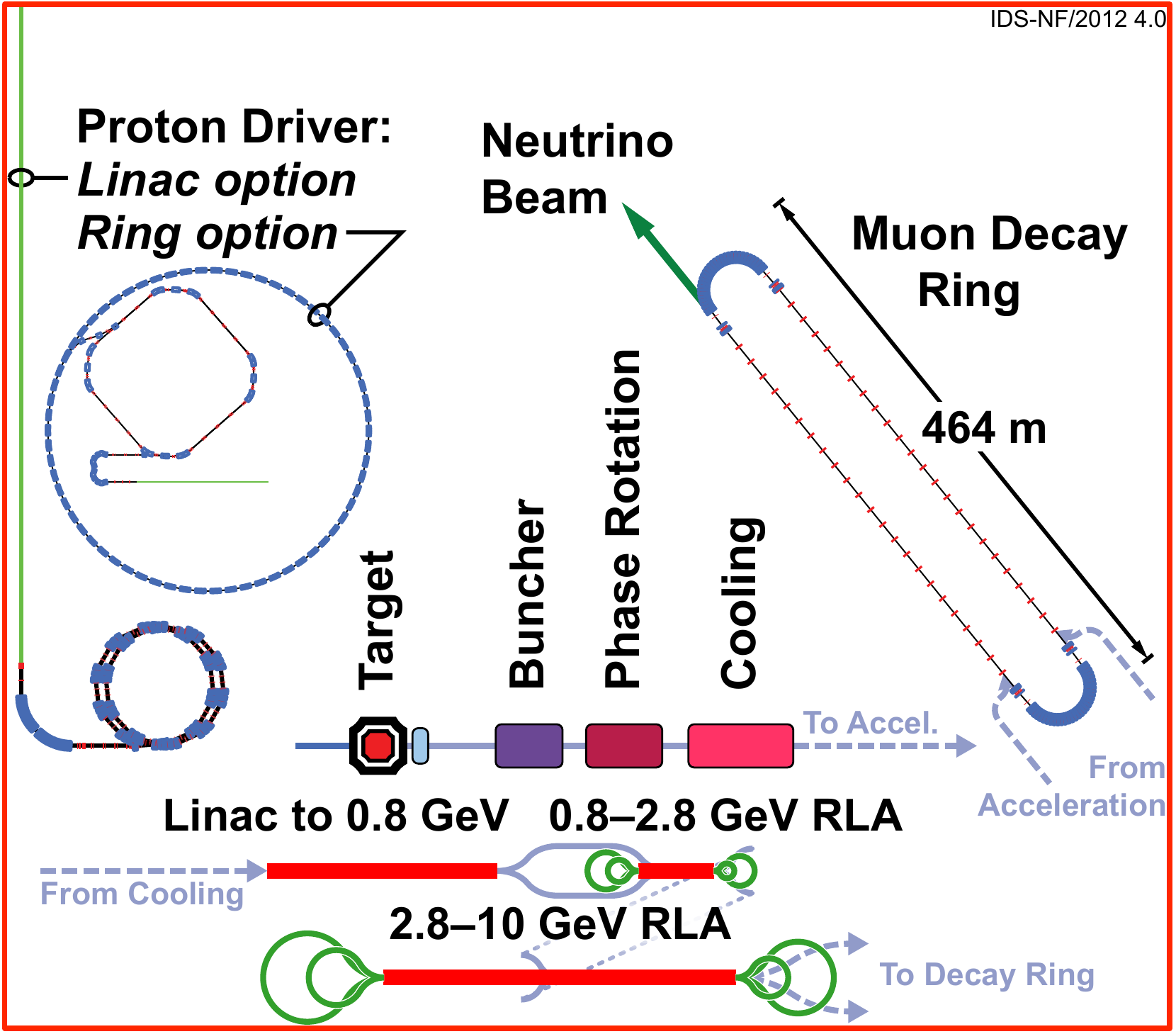}\end{minipage}
\hfill
\begin{minipage}[b]{13pc}
\caption{Two neutrino factory block diagrams: (upper) MAP and (lower) IDS-NF. They differ primarily in the final muon energy.}\label{fig:NF}
\end{minipage}
\end{figure}

Figure~\ref{fig:NF} shows block diagrams of the two
approaches  currently under investigation. The IDS-NF~\cite{IDS-NF,Edgecock}  ``green field" study  finds an optimal baseline of $\approx$\,2,000\,km, matched to a 10\,GeV muon energy and a magnetized-iron detector. The Muon Accelerator Program (MAP)~\cite{MAP} approach starts with the 1,300\,km Fermilab--Homestake baseline, for which the optimal muon energy is 5\,GeV and the required muon-detection momentum threshold correspondingly lower, with the use of a LAr detector (preferably in a magnetic field so as to distinguish detected $\mu^+$ from $\mu^-$) being beneficial. The two approaches have been shown to have comparable physics reach~\cite{Coloma}. They are seen to have much in common,  with differences stemming from the site-specific nature of the MAP design. In particular, they employ the same targetry and cooling approaches.

\section{Targetry}
The neutrino factory goal of $10^{21}$ neutrinos per year aimed at a remote detector requires use of a multi-megawatt proton beam for pion production. Designs exist for beams of the requisite intensity at CERN, Fermilab, and RAL~\cite{Europe,Project-X}. Conventional solid targets are however expected to display inconveniently short lifetimes under such intense proton bombardment~\cite{Densham}, motivating the development of a recirculating liquid jet target. At the relatively low proton energies under consideration (3--8\,GeV), high-$Z$ target materials such as Hg or Pb--Bi eutectic are attractive in order to roughly equalize the $\pi^+$ and $\pi^-$  yields and enable the use of a shorter target, although 
Ga also has advantages~\cite{Edgecock}.
The principle of a liquid-Hg jet was demonstrated in the MERIT experiment at CERN in 2007, which targeted intense proton pulses from the PS on a mercury jet within a 15\,T solenoid at approximately one pulse per hour. Instantaneous pulse intensities consistent with 8\,MW beam power at a 70\,Hz pulse repetition rate were shown to be feasible~\cite{McDonald}.

\begin{figure}
\includegraphics[width=\linewidth,trim=25 10 60 10,clip]{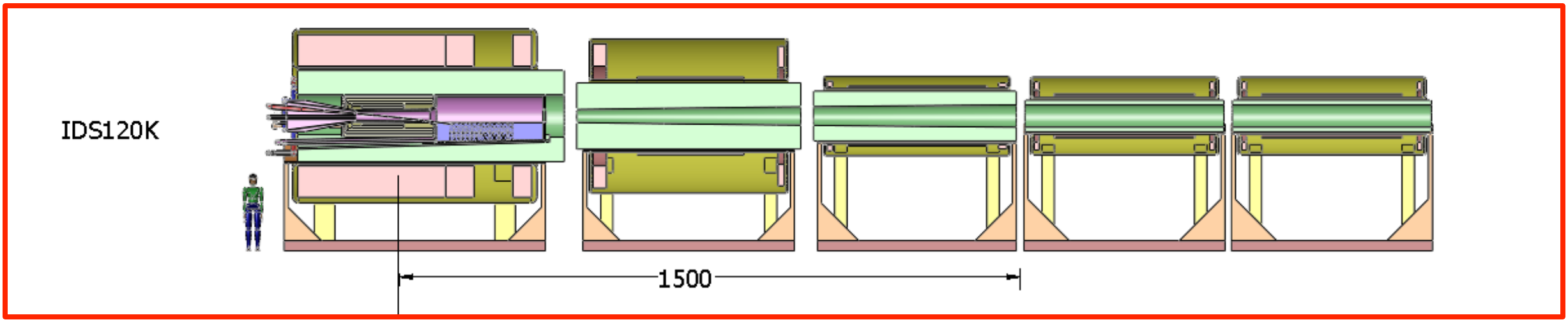}
\includegraphics[width=.5\linewidth,trim=5 5 5 5,clip
]{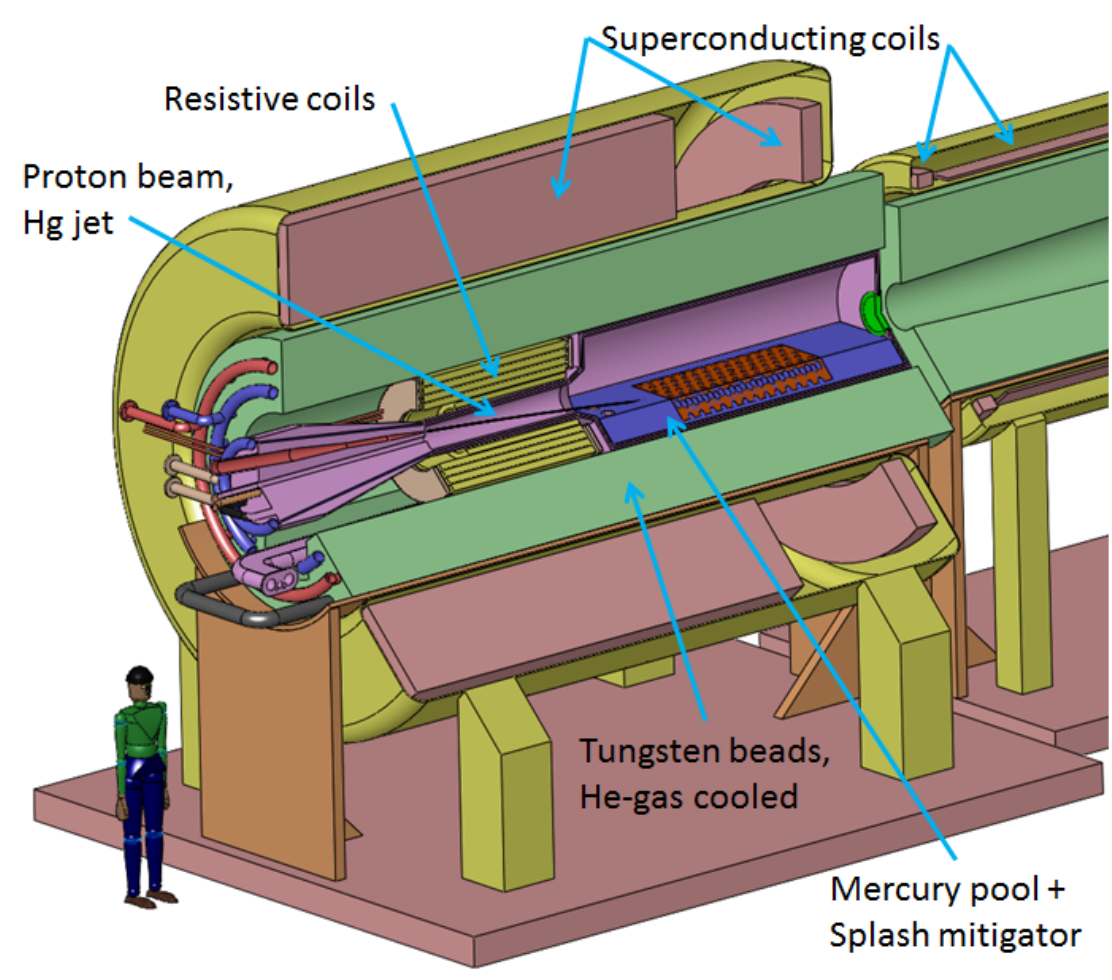}\hfill
\begin{minipage}[b]{17pc}
\caption{Schematic diagrams of mercury-jet target and capture solenoids in the IDS-NF neutrino factory design, specified to handle a 4\,MW proton-beam power at 5--15\,GeV beam energy}\label{fig:tgt}
\end{minipage}
\end{figure}

Current development work~\cite{Tgt-Capture} centers on the magnet and shielding arrangement needed in order to capture pions produced in the target, and their decay muons, with high efficiency, while protecting the superconducting coils from  beam-induced radiation. Figure~\ref{fig:tgt} shows a recent design. It incorporates a 15--20\,T hybrid solenoid magnet, comprising a superconducting outsert and copper insert, surrounding the target. Proceeding downstream, the mercury collection pool also serves as the beam dump, and the solenoid field tapers gradually to 1.5--2\,T, the exact field values and profile being the subject of ongoing optimization studies. Copious amounts of  shielding are deployed, necessarily increasing the solenoid bore and stored energy. The initial capture solenoid has a $\approx$\,3\,GJ stored energy\,---\,very large by accelerator standards, but small compared, e.g., to that of ITER. Alternative designs using high-temperature superconductor may have advantages and are being explored. 

Given the multiple projects worldwide that are planned for multi-megawatt beams, engineering solutions are expected to be available by the time they are needed. A neutrino factory built in advance of such solutions is nevertheless worthwhile since useful sensitivity beyond that available at superbeams is achievable with intensity a factor $\approx$\,6--10 lower than that of the ``ultimate" (e.g., IDS-NF) facility~\cite{Coloma2,Edgecock}. This could be achieved (for example) by starting out at 1\,MW proton-beam power (as discussed for superbeams) and postponing muon cooling (discussed next) to a later upgrade. This is in fact the approach advocated by MAP~\cite{MASS}.

\section{Cooling}
Muon cooling, while not essential for a neutrino factory, has been shown cost-effective as a means of increasing the rate of detected events, as compared to putting more beam on target or substantially increasing the size of the detector. Only ionization cooling~\cite{cooling1} is fast enough to cool muons before a large fraction of them have decayed. It uses energy loss in low-$Z$ absorbers to reduce the divergence of the beam; alternating-gradient focusing brings a concomitant reduction of the beam's cross-sectional area. The competing process is multiple Coulomb scattering, thus the rate of normalized-emittance change per unit absorber length is given by~\cite{cooling2}
\begin{eqnarray} 
\frac{d\epsilon_n}{ds}\approx
-\frac{1}{\beta^2} \left\langle\!\frac{dE_{\mu}}{ds}\!\!\right\rangle\frac{\epsilon_n}{E_{\mu}}
 +
\frac{1}{\beta^3} \frac{\beta_\perp
(0.014\,{\rm GeV})^2}{2E_{\mu}m_{\mu}L_R}\,,
\label{eq:cool} 
\end{eqnarray}  
where  $\beta c$,  $E_\mu$,  and $m_\mu$ are
the muon velocity, energy, and mass; $\beta_\perp$ the lattice betatron function (focal length) at the absorber location; and $L_R$ the  radiation length of the absorber material. Low-$Z$ absorber media are thus preferred, with LiH the medium of choice in recent NF studies. 

\begin{figure}
\centerline{\includegraphics[width=.45\linewidth,trim=17 10 20 10,clip]{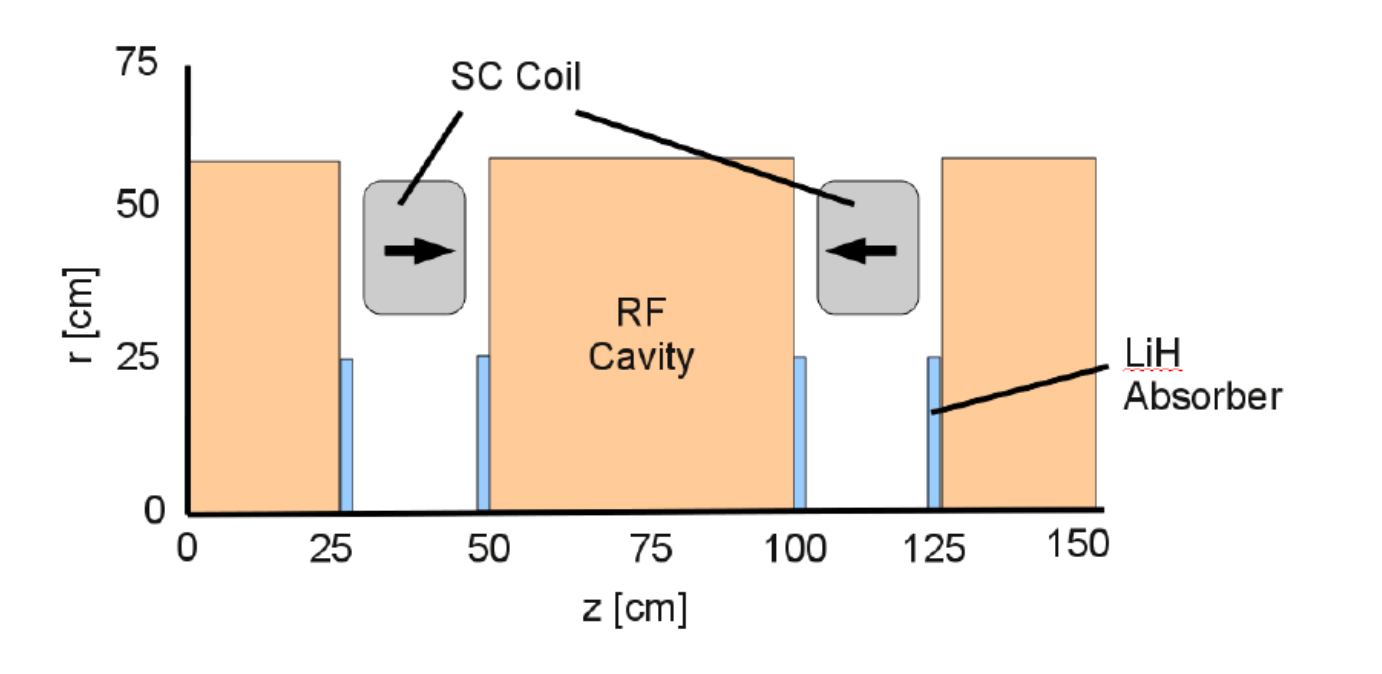}\includegraphics[width=.233\linewidth,trim=0 5 107.7 25,clip
]{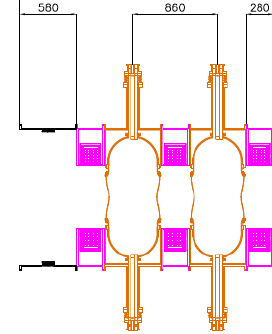}\includegraphics[width=.2338\linewidth,trim=102.5 5 5 25,clip
]{IDS-cool-eng}}
\caption{Concept and engineering diagrams of muon cooling channel in the IDS-NF~\cite{IDS-NF} neutrino factory design, employing LiH absorbers and 201.25\,MHz RF cavities.
}
\label{fig:cool}
\end{figure}

Figure~\ref{fig:cool} shows the cooling channel used in the IDS-NF study~\cite{IDS-NF}, which delivers a factor $\approx$\,2.2 in stored-muon intensity.
Essentially, it is a linac with low-$Z$ absorber material inserted. To be efficient, as much of the lattice as possible should contain absorber. However, the energy-loss rates available from low-$Z$ absorbers ($\sim$\,100\,MeV/m) substantially exceed  ``real-estate" RF accelerating gradients ($\sim$\,10\,MV/m), thus in practice the RF cavities dominate the length of the channel. To achieve low $\beta_\perp$ within the shortest possible distance, solenoids are preferred over the more conventional, quadrupole focusing. 

While none of the ingredients of a muon cooling channel is new, their combination in the close confines of a lattice that is as short as possible  is novel. It has therefore been deemed important to demonstrate experimentally the feasibility of building such a channel and operating it in a muon beam; this also affords the opportunity to test in detail the physics models used in designing ionization-cooling channels. The result is the international Muon Ionization Cooling Experiment (MICE)~\cite{MICE}, under construction at the UK's Rutherford Appleton Laboratory (discussed elsewhere in these Proceedings~\cite{Adey}).

\subsection{Normal-Conducting RF Cavity R\&D}
Ionization cooling's large needed RF-bucket area and race against muon decay place high-gradient RF cavities in strong solenoidal magnetic fields. R\&D on such cavities has shown that strong solenoidal focusing exacerbates breakdown~\cite{Norem}, potentially limiting the RF gradient that can be achieved. Cooling-channel R\&D thus includes an effort to understand and mitigate the sources of such breakdown~\cite{Torun}. Recent progress includes the demonstration that a high-pressure gaseous hydrogen fill can totally suppress magnetic-field-induced breakdown, and that the resulting cavity plasma loading (due to ionization electrons) can be managed by introduction of a sub-percent admixture of electronegative impurity gas~\cite{Chung}. Hydrogen-pressurized cooling channels thus appear feasible, though they do present unique engineering challenges~\cite{Yonehara}. Promising ideas to suppress background in vacuum cavities are also under investigation~\cite{Torun}.

\section{Other challenges}
Although they have novel elements, the following key sections of a neutrino factory are more straightforward than the cooling and targetry.
\subsection{Proton Driver}
As mentioned, multiple multi-MW proton sources have been discussed, and a number are in various stages of design, upgrading, or construction around the world~\cite{Project-X,ESS}. The MAP plan is based on Project X at Fermilab, whose reference design~\cite{Project-X} envisions a  CW 1\,GeV, 1\,MW H$^-$ linac feeding, in turn,  1--3\,GeV, 3\,MW and  3--8\,GeV, 0.3\,MW pulsed linacs. Accumulator and compressor rings are added (Fig.~\ref{fig:NF}) in order to provide a proton beam on target with the requisite duty factor. Useful neutrino factory sensitivity is already achievable with 1\,MW at 3\,GeV (``NuMAX"), which can later be upgraded  (to ``NuMAX+") by the addition of a 3\,MW target and muon-cooling channel~\cite{MASS}. A further upgrade scenario has been studied in which the beam current and bunch length in these linacs are increased in order to provide a 4\,MW beam power at 8\,GeV, suitable for a muon collider. The IDS-NF plan envisions either a multi-GeV linac or a 180\,MeV H$^-$ linac feeding two rapid-cycling synchrotrons or FFAGs.

\subsection{Front End}
Following the Target and Capture section of the neutrino factory, the muon beam must be prepared for cooling and acceleration. Pions, and hence, their decay muons, are produced over a wide range of momentum. A drift section employing a solenoid focusing lattice allows a momentum--time correlation to develop, following which the muons are bunched and ``phase rotated" by accelerating the slower ones and decelerating the faster ones. This is accomplished in a ``vernier" RF scheme employing a range of frequencies from (in the IDS-NF version) 320 to 202\,MHz. Simulations have shown that large numbers of higher-energy protons produced in the target, which would cause losses in absorbers and windows, would be transported by such a lattice; to avoid this, they are separated from the muons via a chicane and absorbed in shielding.

\subsection{Acceleration}
\begin{figure}
\includegraphics[width=.7\linewidth,trim=0 0 0 0 mm,clip]{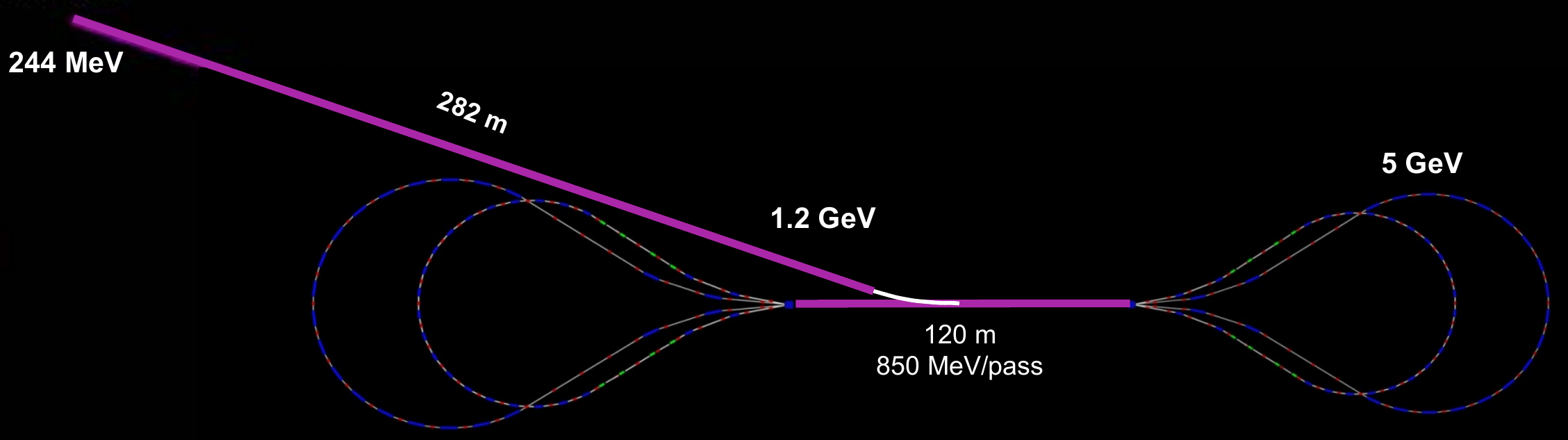}\hfill
\begin{minipage}[b]{10pc}
\caption{MAP neutrino factory muon acceleration design.}
\label{fig:RLA} 
\end{minipage}
\end{figure}

Following the cooling section the muons are accelerated in a linac feeding one or two RLAs (depending on the desired final energy). Figure~\ref{fig:RLA} shows the 5\,GeV MAP design. A previous 25\,GeV design incorporated an FFAG stage as well, but given the measured value of $\theta_{13}$, so high an energy is no longer essential.

\subsection{Decay Ring}
The decay ring must accommodate large numbers of decaying muons and thus must be designed to handle energy deposition from decay electrons. This requires the incorporation of a tungsten liner. A 10\,GeV design has been devised for IDS-NF and can be scaled to 5\,GeV.

\section{Conclusions}
An ``entry level" neutrino factory (NuMAX), without cooling and with a 1\,MW target, has a physics reach exceeding that of a MW-scale superbeam. Given its inherent upgradability (to NuMAX+), it is arguably the preferable option. With only a few open R\&D issues remaining, and a clear path to their resolution, neutrino factory design is reaching the point where the start of a construction project could be envisaged this decade. 

\section*{Acknowledgments}
I thank the organizers for their gracious hospitality and the invitation to present this talk. Work supported by  U.S. Dept.\ of Energy (through MAP) and National Science Foundation.


\section*{References}
\begin{thebibliography}{9}
\bibitem{Geer}
Geer S 1998 {\it Phys.\ Rev.\ D} {\bf 57} 6989

\bibitem{beta}
The alternative ``beta beam" approach is no longer being pursued, having been judged to possess no competitive advantage over a neutrino factory (see Ref.~\protect\cite{Edgecock}).

\bibitem{theta13}
Abe Y {\it et al}\, 2012 {\it Phys Rev D} {\bf 86}, 052008;
An F P {\it et al}\, 2012 {\it Phys Rev Lett}  {\bf 108}, 171803;\\
Ahn J K {\it et al}\, 2012 {\it Phys Rev Lett} {\bf 108}, 191802; 
An F P {\it et al}\, 2013 {\it Chin Phys C} {\bf 37}, 011001

\bibitem{Blennowetal}
Blennow M, Coloma P, Huber P, and Schwetz T 2013 Quantifying the sensitivity of oscillation experiments to the neutrino mass ordering
 {\it Preprint} arXiv:1311.1822 [hep-ph]
 
\bibitem{MASS}
ed Delahaye J-P {\it et al}\, 2013 Enabling Intensity and Energy Frontier Science with a Muon Accelerator Facility in the U.S.:
A White Paper Submitted to the 2013 U.S. Community Summer Study of the Division of Particles and Fields  of the American Physical Society
 {\it Preprint}
FERMILAB-CONF-13-307-APC,	arXiv:1308.0494 [physics.acc-ph], and Delahaye J-P {\it et al}\, contribution to this Workshop

\bibitem{FS}
ed Autin B, Blondel A, and Ellis J 1999 Prospective Study of Muon Storage Rings at CERN {\it Report} CERN 99-02, ECFA 99-197\\
ed Holtkamp N and Finley D 2000 Feasibility Study of a Neutrino Source Based on a Muon Storage Ring  {\it Report}  FERMILAB-PUB-00/108-E\\
ed Ozaki S {\it et al}\, 2001 Feasibility Study-II of a Muon-Based Neutrino Source {\it Report}  BNL-52623\\ 
ed Kuno Y and Mori Y 2003 A Feasibility Study of a Neutrino Factory in Japan  {\it Report} KEK Report \\
Abrams R J {\it et al}\, 2011 International Design Study for the Neutrino Factory, Interim Design Report {\it Preprint} arXiv:1112.2853 [hep-ex]

\bibitem{MAP}
Palmer M A 2013 this Workshop

\bibitem{IDS-NF}
See {\tt https://www.ids-nf.org/wiki/FrontPage}\\
Soler P 2013 this Workshop\\
EUROnu Special Collection 2013 {\it Phys. Rev. ST Accel. Beams}, {\tt http://prst-ab.aps.org/speced/EURONU}

\bibitem{Edgecock}
Edgecock T R {\it et al}\, 2013 High intensity neutrino oscillation facilities in Europe {\it Phys. Rev. ST Accel. Beams} {\bf 16} 021002

\bibitem{Coloma}
Coloma P, Donini A, Fern\'{a}ndez-Mart\'{i}nez E, and Hern\'{a}ndez P 2012 Precision on leptonic mixing parameters at future neutrino oscillation experiments {\it JHEP} {\bf  1206}, 073 

\bibitem{Europe}
Thomason J W G {\it et al}\, 2013 Proton driver scenarios at CERN and Rutherford Appleton Laboratory {\it Phys. Rev. ST Accel. Beams} {\bf 16}, 054801

\bibitem{Project-X}
See {\tt http://projectx.fnal.gov/}, and \\
ed Holmes S D 2013 Project X
Accelerator Reference Design, Physics Opportunities, Broader Impacts {\it Report} arXiv:1306.5022

\bibitem{Densham}
See e.g. Densham C 2013 this Workshop

\bibitem{McDonald}
McDonald K T {\it et al} 2010 The MERIT High-Power Target Experiment at the CERN PS {\it Proc.
 IPAC'10}, WEPE078

\bibitem{Tgt-Capture}
McDonald D 2013 this Workshop;
Sayed H 2013 this Workshop

\bibitem{Coloma2}
Christensen E, Coloma P, and Huber P 2013 Physics Performance of a Low-Luminosity Low Energy Neutrino Factory {\it Preprint} arXiv:1301.7727 [hep-ph]

\bibitem{cooling1}
Ado Y M and  Balbekov V I 1971 Use of ionization friction in the storage of heavy particles {\it At.\ Energ.}\  31(1) 40 (1971) 
{\tt http://www.springerlink.com/content/v766810126338571/}; see also 
Neuffer D  1987 {\it AIP Conf.\ Proc.}\ {\bf 156}, 201 and references therein, and
Fernow R C and Gallardo  J C 1995 {\it Phys.\ Rev.\ E} {\bf 52}, 1039

\bibitem{cooling2}
Neuffer D  1999 $\mu^+\mu^-$ Colliders {\it  Report} CERN-99-12

\bibitem{MICE}
See {\tt http://mice.iit.edu/}\\
Gregoire G {\it et al}\, 2003 Proposal to the Rutherford Appleton Laboratory: An International Muon Ionization Cooling Experiment (MICE) {\tt http://mice.iit.edu/micenotes/public/pdf/MICE0021/MICE0021.pdf}

\bibitem{Adey}
Adey D 2013 this Workshop;
Kaplan D M 2013 this Workshop

\bibitem{Norem}
Norem J {\it et al}\, Dark Current, Breakdown, and Magnetic Field Effects in a Multicell, 805 MHz Cavity 2003 {\it Phys.\ Rev.\ ST Accel. Beams} {\bf 6}, 072001

\bibitem{Torun}
Torun Y {\it et al}\, 2011 MuCool R\&D {\it ICFA Beam Dynamics Newslett.} {\bf 55}, 103\\
Bowring D {\it et al}\, 2013 A Modular Cavity for Muon Ionization Cooling R\&D
{\it Proc. IPAC2013}, WEPFI073

\bibitem{Chung}
Chung M {\it et al}\, 2013 Pressurized H$_2$ RF Cavities in Ionizing Beams and Magnetic Fields {\it Phys.\ Rev.\ Lett.}\ {\bf 111}, 184802\\
Freemire B 2013 High Pressure Gas Filled RF Cavity Beam Test at the Fermilab MuCool Test Area
{\it PhD thesis} Illinois Institute of Technology

\bibitem{Yonehara}
Yonehara K 2013 this Workshop

\bibitem{ESS}
Dracos M 2013 this Workshop

\end{thebibliography}
\end{document}